\documentstyle[aps,epsf,epsfig,prl,multicol]{revtex}
\begin{document}

\vspace{0.0cm}
\draft

\title{
Entropy production and wave packet dynamics in the Fock space of
closed chaotic many-body systems. }
\author{V.V. Flambaum $^1$ \thanks{email address:
flambaum@newt.phys.unsw.edu.au} and F.M. Izrailev $^2$}

\address{$^1$ School of Physics, University of New South Wales,
Sydney 2052, Australia\\ $^2$Instituto de F\'{\i}sica, Universidad
Aut\'onoma de Puebla, Apartado Postal J-48, Puebla 72570,
M\'exico}

\maketitle

\begin{abstract}
Highly excited many-particle states in quantum systems such as nuclei,
atoms, quantum dots, spin systems, quantum computers etc., can be considered
as ``chaotic'' superpositions of mean-field basis states (Slater
determinants, products of spin or qubit states). This is due to a very high
level density of many-body states that are easily mixed by a residual
interaction between particles (quasi-particles). For such systems, we have
derived simple analytical expressions for the time dependence of energy
width of wave packets, as well as for the entropy, number of principal basis
components and inverse participation ratio, and tested them in numerical
experiments. It is shown that the energy width $\Delta (t)$ increases
linearly and very quickly saturates. The entropy of a system increases
quadratically, $S(t) \sim t^2$ at small times, and after, can grow linearly,
$S(t) \sim t$, before the saturation. Correspondingly, the number of
principal components determined by the entropy, $N_{pc} \sim exp{(S(t))}$,
or by the inverse participation ratio, increases exponentially fast before
the saturation. These results are explained in terms of a cascade model
which describes the flow of excitation in the Fock space of basis
components. Finally, a striking phenomenon of damped oscillations in the
Fock space at the transition to an equilibrium is discussed.
\end{abstract}
\date{21.03.2001}
\pacs{PACS numbers:  03.67.Lx, 05.45.Mt, 24.10.Cn}
\begin{multicols}{2}

\section{Introduction}

Highly excited many-particle states in many-body systems quite often can be
presented as ``chaotic'' superpositions of shell-model basis states, see
recent calculations for complex atoms \cite{FGGK94}, multicharged ions \cite
{ions}, nuclei \cite{nuclei} and spin systems \cite{Nobel,spins}. The origin
of this phenomenon relates to a very high density of many-particle energy
levels, which increases drastically with an increase of energy. Indeed, the
number $N$ of combinations in the distribution of $n$ particles (or
quasi-particles) over $m$ ``orbitals'' (single-particle states) is
exponentially large ($N\sim m!/n!(m-n)!$ in a Fermi system). Therefore, the
spacing $D$ between many-body levels is exponentially small and a
``residual'' interaction $V$ between the particles can mix a huge number of
the basis states of a mean field $H_0$ (Slater determinants), when forming
exact eigenstates of the total Hamiltonian $H=H_0+V$.

The onset of chaos for highly excited states, as well as for many-particle
spectra, has been recently studied in great detail in terms of the Two-Body
Random Interaction (TBRI) model which was invented about three decades ago
\cite{old}. In this model all {\it two-body} matrix elements are assumed to
be independent and random variables, therefore all dynamical correlations
are neglected. Thus, the TBRI model is essentially the random matrix model,
however, it differs from standard Random Matrix Models where the two-body
nature of interaction is not taken into account (see, e.g. \cite
{FGI96,FI97,I00v} and review \cite{GMW99}).

One of the important results obtained recently \cite{AGKL97} in the frame of
this model is the Anderson-like transition which occurs in the Fock space
determined by many-particle states of $H_0$. The critical value $V_{cr}$ for
this transition is determined by the density of states $\rho_f=d_f^{-1}$ of
those basis states which are directly coupled by a two-body interaction \cite
{FGI96,FI97,AGKL97,all,FI00}. When the interaction is very weak, $V_0\ll d_f$
, exact eigenstates are delta-like functions in the unperturbed basis, with
a very small admixture of other components which can be found by the
standard perturbation theory. With an increase of the interaction, the
number of principal components $N_{pc}$ increases and can be very large, $%
N_{pc}\gg 1$. However, if the interaction is still not too strong, $\pi^{-2}
\sqrt{d_fD}\ll V_0 \leq d_f$ \cite{FI97}\thinspace , the eigenstates are
sparse, with extremely large fluctuations of components. In order to have
ergodic eigenstates which fill some energy range (see below), one needs to
have the perturbation large enough, $V_0\gg d_f\,$(for a large number of
particles this transition is sharp and, in fact, one needs the weaker
condition, $V_0\geq d_f$ ).

Above the threshold of chaos, $V_0\geq d_f$, the number of principal basis
components in an eigenstate can be estimated as $N_{pc} \sim\Gamma/D$ where $%
\Gamma$ is the spreading width of the {\it strength function}. In such
chaotic eigenstates any external weak perturbation is exponentially
enhanced. The enhancement factor can be estimated as $\sqrt{N_{pc}} \propto
1/\sqrt{D}$, see e.g. \cite{enhancement} and references therein. This huge
enhancement have been observed in numerous experiments when studying parity
violation effects in compound nuclei (see, for example, review \cite{W}).

In recent works \cite{S,F2000} the theory of many-body chaos has been
extended to quantum computers. Since in this case the density of energy
levels is extremely high, it is often impossible to resolve particular
many-body levels. This happens for the injection of an electron into a
many-electron quantum dot, for the capture of an energetic particle by a
nucleus or atom, or for different models of quantum computer with a large
number of interacting qubits (spins). In this case the approach based on the
study of stationary chaotic eigenstates turns out to be not an adequate one,
and one should consider the time evolution of a wave function and entropy
\cite{F2000}.

In contrast to the study of spectra and eigenstates, the analysis of the
evolution of wave packets in random matrix models is based mainly on
numerical results. In the first line, one should mention the numerical study
\cite{IKPRT97} of Band Random Matrices that describe quasi-1D disordered
models with finite number of channels. Recently, the attention has been paid
to the so-called Wigner Band Random (WBRM) model \cite{W55} which is used in
the study of generic properties of strength functions in dependence on the
strength of interaction \cite{CCGI96,FCIC96}. In particular, in Refs.\cite
{CIK00,doron} the problem of the quantum-classical correspondence for the
time evolution of wave packets was under close study. We note that the WBRM
model serves as convenient random matrix model for different quantum
systems, and in many aspects can be compared with the TBRI model \cite{I00n}.

In this paper we study generic properties of the evolution of wave packets
in the energy shell, paying the main attention to the time-dependence of the
entropy, width of packets, and inverse participation ratio. We derive
analytical estimates for these quantities and check numerically our
predictions with the use of the TBRI and WBRM models.


\section{Time evolution of chaotic many-body states}

Exact (``compound'') eigenstates $\left| k\right\rangle \,$ of a many-body
Hamiltonian $H=H_0+V$ can be expressed in terms of simple shell-model basis
states $\left| f\right\rangle $ (eigenstates of $H_0$), or by products of
qubits in a model of quantum computer:
\begin{equation}
\label{slat}\left| k\right\rangle =\sum\limits_f C_f^{(k)}\left|
f\right\rangle \,;\,\,\,\,\,\,\,\,\left| f\right\rangle
=a_{f_1}^{+}...a_{f_n}^{+}\left|^{\prime\prime}vacuum^{\prime\prime}\right
\rangle.
\end{equation}
Here $a_s^{+}$ are creation, or spin-raising, operators (in the latter case,
the ground state $\left|^{\prime\prime}vacuum ^{\prime\prime}\right\rangle$
corresponds to the situation with all spins down), and $C_f^{(k)}$ are
components of compound eigenstates $\left| k\right\rangle $ formed by the
residual interaction $V$.

In what follows, we consider the time evolution of the system for the case
when compound eigenstates are {\it chaotic}. By this term we mean that the
number of principal components is very large, $\sqrt{N_{pc}}\gg 1$, and the
components $C_f^{(k)}$ can be treated as uncorrelated amplitudes with the
gaussian distribution around their mean values, see details in \cite{FI97}.
Let us assume that initially ($t=0$) the system is in a specific basis state
$\left| 0\right\rangle $ (with certain orbitals occupied, or, in the case of
a quantum computer, when the state with certain spins ``up'' is prepared).
This state can be presented as a sum over exact eigenstates of the total
Hamiltonian $H$:
\begin{equation}
\label{in}\left| 0\right\rangle =\sum\limits_kC_0^{(k)}\left| k\right\rangle
.
\end{equation}
Then the time-dependent wave function reads as
\begin{equation}
\label{psit}\Psi (t)=\sum\limits_{k,f}C_0^{(k)}C_f^{(k)}\left|
f\right\rangle \exp (-iE^{(k)}t).
\end{equation}
Here the sum is taken over the compound states $k$ and basis states $f$, and
we set $\hbar =1$.

The probability $W_0=|A_0|^2 =|\left\langle 0|\Psi(t)\right\rangle|^2$ to
find the system in the initial state is determined by the amplitude
\begin{eqnarray}
\label{ampli}
A_0= \left\langle 0|\exp(-iHt)|0\right\rangle=
\sum\limits_k|C_0^{(k)}|^2\exp(-i E^{(k)}t) \simeq
\nonumber \\
\int dE P_0(E)\exp(-i Et).
\end{eqnarray}
Here we replaced the summation over a large number of the eigenstates by the
integration over their energies $E \equiv E^{(k)}$, and introduced the {\it %
strength function} (SF) $P_0(E)$ which is also known in the literature as
the {\it local spectral density of states}, LDOS,
\begin{equation}
\label{strength}P_0(E)\equiv \overline{|C_0^{(k)}|^2}\rho (E),
\end{equation}
with $\rho (E)$ as the density of exact eigenstates.

As one can see, the probability $W_0$ is entirely determined by the strength
function (\ref{strength}). It is well known that in many applications this
function has the Breit-Wigner (BW) form with a half-width $\Gamma_0=2\pi\rho
H_f^2$. In our case $\rho=\rho_f$ is the density of directly coupled basis
states and $H_f^2=\overline{\left| H_{0f}\right| ^2}$ is the variance of the
non-zero off-diagonal elements of $H$, defined by the residual interaction $V
$. One should remind that in real situations the second moment $\Delta^2_E$
of the SF is always finite due to a finite range of interaction in the
energy representation. Therefore, the Breit-Wigner form of the SF can occur
for a finite energy range only, determined by the energy width of the
interaction.

However, it was recently shown (see e.g. \cite{FI97,FI00} and references
therein) that if $\Gamma_0$ defined by the above expression, is of the order
(or larger) of the mean-square-root width $\Delta_E$ of the SF itself, the
form of the SF in the TBRI model is very close to the Gaussian. Strong
deviations of the SF from the BW-dependence have been observed numerically
when studying the structure of the SF and eigenfunctions of the Ce atom \cite
{FGGK94}. Also, numerical data \cite{nuclei} have revealed that the form of
the SF in nuclear shell models is much closer to the Gaussian rather that to
the BW. This results from the fact that the three orbitals $s$, $d_{3/2}$
and $d_{5/2}$ included into the nuclear shell-model calculation \cite{nuclei}%
, have close mean-field energies and the residual interaction $V$ plays the
dominant role in the Hamiltonian matrix.

Recent analytical results \cite{FI00} for the TBRI model allow to describe
the whole transition for the SF from the BW regime to that of the Gaussian.
This model is characterized by two-body random matrix elements which
determine the residual interaction $V$ between $n$ Fermi-particles occupying
$m$ orbitals (single-particle states), see details and references, for
example, in \cite{FI97,I00n}. It was shown that in general case the SF is
given by the following approximate expression \cite{BM69,FI00},
\begin{equation}
\label{FfBW}P_0(E)=\frac 1{2\pi }\frac{\Gamma (E)}{(E_0+\delta _0 (E)
-E)^2+\Gamma ^2(E)/4},
\end{equation}
\begin{equation}
\label{GammaH}\Gamma (E)\simeq 2\pi \overline{\left| H_{0f}\right| ^2}\rho
_f(E).
\end{equation}
One can see that the general expression (\ref{FfBW}) is of the Breit-Wigner
form, however, with $\Gamma (E)$ as some function of the total energy $E$.
Here $\delta _0 (E)$ is the correction to the unperturbed energy level $E_0$
due to the residual interaction $V$, and $\rho _f(E)$ is the density of
those basis states $\left| f\right\rangle $ which are directly coupled by
the interaction $H_{0f}$ with the initial state $\left| 0\right\rangle $. It
was shown \cite{FI00} that for the TBRI model the function $\Gamma (E)$ has
the Gaussian form with the variance which depends on the model parameters.
In the case of relatively small (but non-perturbative) interaction, $\Gamma
_0=2\pi \rho _f H_f^2\ll \Delta _E$, the function $\Gamma (E)$ is very broad
(i.e. it does not change significantly within the energy intervals $\sim
\Gamma$ and $\Delta _E$) and can be treated as the constant, $\Gamma (E)
\simeq \Gamma _0$. In the other limit case of a strong interaction, $\Gamma
_0\ge \Delta _E$, the dependence $\Gamma (E)$ in (\ref{FfBW}) is the leading
one, and the slow dependence in the denominator can be neglected. One should
note that the simple expression (\ref{GammaH}) for the width $\Gamma (E)$
has to be modified in this limit, see details in \cite{FI00}.

The second moment $\Delta^2_E$ of the SF can be found from the equation $%
\Delta_E^2= \sum\limits_{f \neq 0} H_{0f}^2$. Here the summation is taken
over the off-diagonal matrix elements $H_{0f}$ which couple the initial
state $\left| 0\right\rangle$ with the others $\left| f\right\rangle$. For
the TBRI model the analytical expression for $\Delta_E$ has been obtained in
\cite{FI97},
\begin{equation}
\label{SFwidth} \Delta_E^2=\frac{1}{4} V_0^2 n(n-1)(m-n)(m-n+3)
\end{equation}
where $V_0^2$ is the variance of the off-diagonal matrix elements of the
two-body residual interaction $V$. In fact, for Fermi-particles the width $%
\Delta_E$ turns out to be the same for any basis state $\left|0 \right
\rangle$.

Let us first start with the probability $W_0(t)$ of the system to stay in
the initial state. In two limit cases of small and very large times, the
dependence $W_0$ is shown \cite{F2000,FI00b} to be of the following forms,
\begin{equation}
\label{wgauss}W_0(t)=\exp \left( -\Delta _E^2t^2\right)
;\,\,\,\,\,\,\,\,\,\,\,\,\,t\ll \frac{\Gamma _p}{\Delta _E^2}
\end{equation}
and
\begin{equation}
\label{wBW}W_0(t)=C\exp \left( -\Gamma _p\,t\right)
;\,\,\,\,\,\,\,\,\,\,\,\,\,\,\,\,\,t\gg \frac{\Gamma _p}{\Delta _E^2}\,\,.
\end{equation}
Here $\Gamma _p$ is the imaginary part of the pole of the SF (see Eq.(\ref
{FfBW}) in the complex energy plane. In the case when the SF has the
standard BW form, we have the obvious relation $\Gamma _p=\Gamma _0$ where
the latter is given by the Fermi golden rule. In other limit of a strong
interaction, when the SF has the Gaussian form, the expression for $\Gamma _p
$ is not simple. The transition from one regime of the time dependence of $%
W_0(t)$ to another is schematically shown in Fig.1.

\begin{figure}[htb]
\vspace{-1.3cm}
\begin{center}
\hspace{-1.5cm}
\epsfig{file=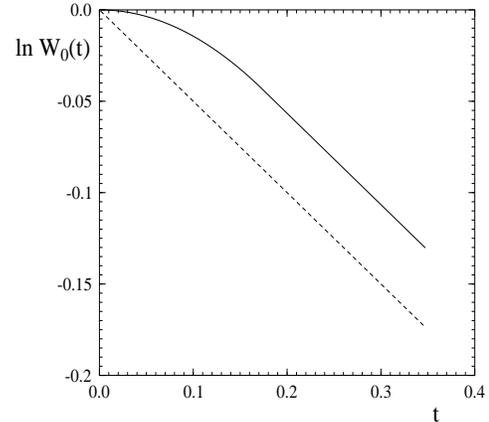,width=3.0in,height=2.3in,angle=-90}
\vspace{-0.8cm}
\caption{
Schematic time dependence of $W_0(t)$ for $\Gamma_p=0.5,
\Delta_E=1.2$; the point $t_c$ where the dependence (\ref{wgauss})
is changed to (\ref{wBW}) is $t_c=\Gamma_p/\Delta_E^2 \approx
0.17$. }
\end{center}
\end{figure}

Now we estimate the probabilities $w_f=|A_f(t)|^2$ to find the system in
other basis states. For a very small time we have,
\begin{equation}
w_f=|\left\langle f|e^{-iHt}|0\right\rangle |^2\simeq |H_{0f}|^2t^2.
\end{equation}
Note, that only the states directly connected to the initial state are
populated at this time scale. One can estimate the population of these
states for a larger time by substituting the time-dependent wave function $%
\Psi (t)=A_0(t)\psi _0+\sum_fA_f(t)\psi _f$ into the Schrodinger equation,
\begin{equation}
\label{Sch}i\hbar \frac{dA_f}{dt}=H_{0f}(t)A_0+\sum_kH_{fk}(t)A_k.
\end{equation}
Here $k,f\neq 0$ and $H_{0f}(t)\equiv H_{0f}\exp (i\omega _{0f}t)$. Note
that the second term in the right-hand-side may be treated as a random
variable with the zero mean value. Indeed, $A_k\propto H_{0k}$, therefore, $
\overline{H_{fk}(t)H_{0f}}=0$. The variance of this term is $\overline{%
\sum_k|H_{fk}(t)A_k|^2}=\overline{|H_{fk}(t)|^2}\sum_k|A_k|^2=\overline{%
|H_{fk}(t)|^2}(1-|A_0|^2)$. Comparing this with the first term in the
right-hand-side of Eq.(\ref{Sch}), one may conclude that the second term is
not very important for small times when $A_0(t)\sim 1$. Neglecting the
second term and assuming $|A_0(t)|=\exp \left( -\Gamma \,t/2\right) $ which
is valid for $\Gamma <<\Delta _E$, we obtain \cite{F2000},
\begin{eqnarray}
\label{Wf}
w_f=
|H_{0f}|^2\left|\int\limits_0^t |A_0(t)| e^{i\omega_{0f}t}
dt\right|^2
\simeq
\nonumber \\
\frac{|H_{0f}|^2}{\omega_{0f}^2+\Gamma^2/4}
\left|e^{(i\omega_{0f}-\Gamma_/2)t}
-1\right|^2 ,
\end{eqnarray}
where $\omega _{0f}=E_f-E_0$. This approximate estimate shows that only the
basis states within the energy interval $\Gamma $ can be substantially
populated (if $\Gamma >\Delta _E$, this energy interval is equal to $\Delta
_E$).

For large time, the result is different for perturbative and chaotic
regimes. In the perturbative regime the expression (\ref{Wf}) for $w_f$ is
the final one. In the chaotic regime the asymptotic expression for $%
t\rightarrow \infty $ can be obtained in the following way. The projection
of $\Psi (t)$ (see Eq. (\ref{psit})) onto the state $f$ gives
\begin{equation}
\label{sfluct}w_f(t)=w_f^s+w_f^{fluct}(t),
\end{equation}
\begin{eqnarray}
\label{ws}
w_f^s =\sum\limits_k|C_0^{(k)}|^2|C_f^{(k)}|^2 \simeq
\int \frac{dE}{\rho(E)} P_0(E) P_f(E) \simeq
\nonumber \\
\frac{1}{2\pi\rho}\frac{\Gamma_t}{(E_0-E_f)^2 +
(\Gamma_t/2)^2} .
\end{eqnarray}
Here the result is written for the case when the SF has the BW form both for
initial $\left| 0\right\rangle $ and final state $\left| f\right\rangle $
with the corresponding half-widths $\Gamma _0^0$ and $\Gamma _0^f$. In this
case the resulting form of $w_f^s$ is, approximately, again the Breit-Wigner
with the new half-width $\Gamma _t\simeq \Gamma _0^0+\Gamma _0^f\simeq
2\Gamma _0$. However, if $\Gamma _0\geq \Delta _E$, the form of $w_f^s$ is
close to the Gaussian with the variance $(\Delta _E)_t^2\simeq 2\Delta _E^2$
\cite{F2000}.

The term $w_f^{fluct}(t)$ can be written in the form,
\begin{equation}
\label{fluc} w_f^{fluct}(t)=\sum\limits_{k,p;k\neq p}C_0^{(k)}C_f^{(k)}
C_0^{(p)}C_f^{(p)} \exp(i(E^{(k)}-E^{(p)})t).
\end{equation}
At large time, $t \rightarrow \infty$, the terms in the sum rapidly
oscillate and one can put $\overline{w_f^{fluct}(t)}=0 $. Thus,
asymptotically the distribution of the components in the time-dependent wave
function is close to that given by the form of the strength function (see
Eqs.(\ref{strength}, \ref{FfBW})), with a slightly larger spreading width.

Note that similar expression for $W_0$ contains the term $|C_0^{(k)}|^4$.
For gaussian fluctuations of the components $C_0^{(k)}$, one can get $
\overline{ |C_0^{(k)}|^4}$ = 3 $(\overline{|C_0^{(k)}|^2})^2$ which is the
known result in the Random Matrix Theory \cite{brody}. Therefore, if the
number of principal components $N_{pc}$ in the SF is very large, the
probability to find the system in the initial state $\left| 0\right\rangle $
at large time is, at least, three times larger than the probability to find
the system in any other state $\left| f\right\rangle $, see Figs. 2,3.

\begin{figure}[htb]
\vspace{-1.5cm}
\begin{center}
\hspace{-1.5cm}
\epsfig{file=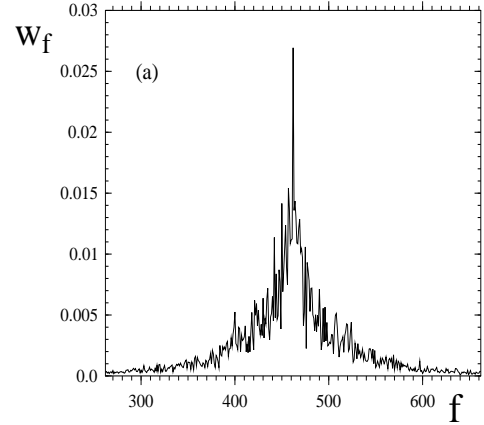,width=3.0in,height=2.3in,angle=-90}
\vspace{-0.8cm}
\caption{
Asymptotic distribution $w_f$ for the case when the strength function is of
the Breit-Wigner form in the TBRI model. The parameters~are:~$%
n=6,\,m=12,\,V_0^2 \approx 0.003$, $\Gamma_0 \approx
0.50,\,\Delta_E \approx 1.16,$ with the average over $N_g=10$
matrices with different realization of random two-body matrix
elements, see in the text. }
\end{center}
\end{figure}

\begin{figure}[htb]
\vspace{-2.5cm}
\begin{center}
\hspace{-1.5cm}
\epsfig{file=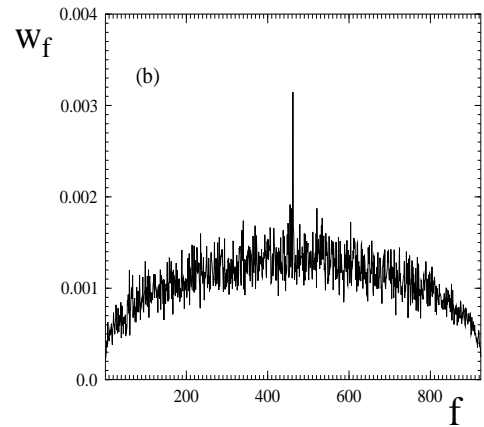,width=3.0in,height=2.3in,angle=-90}
\vspace{-0.8cm}
\caption{
Asymptotic distribution $w_f$ for the case when the strength function is
close to the Gaussian. The only difference from Fig.2 is the interaction
strength, $V_0^2\approx 0.083$, correspondingly, $\Gamma_0\approx 10.5$ and $%
\Delta_E \approx 5.8$; the average was taken over $N_g=50$ matrices. }
\end{center}
\end{figure}

In these figures the distribution of probabilities $w_f$ in the TBRI model
is shown after a very long time $t=40$ for two different strengths of
interaction (in fact, the time $t$ is measured in units $\hbar/d_0$ where $%
d_0=<\epsilon _{s+1}-\epsilon _s>$ is the mean level spacing between
single-particle energies $\epsilon_s$). In both cases $n=6$ Fermi-particles
occupy $m=12$ orbitals, therefore, the total number of many-particle states
(the size of the Hamiltonian matrix) is $N=924 $ . The distribution of $%
\epsilon _s$ is taken random with $d_0=1$. Two-body matrix elements are
taken as gaussian random entries with the zero mean and variance $V_0^2$,
and in order to reduce the fluctuations, the distribution $w_f$ is obtained
with an average over a number $N_g$ of matrices with different random
realizations. Initially, only one basis state $n_0=462$ was populated at the
center of the energy spectrum, in order to avoid the asymmetry of the
distribution in the basis representation.

The two different values of $V_0$, for which the distributions $w_f$ are
obtained, are chosen in such a way that in one case, see Fig.2, the strength
function has the Breit-Wigner form, and in the other, the form is very close
to the Gaussian (Fig.3). We should remind that the above two forms occur in
the energy representation, however, the results are shown in the basis
representation. These two representations are related through the density of
states which is known to be of the Gaussian form for large number of
particles and orbitals \cite{old,brody}.

One can see that in both cases, Fig.2-3, the probability to stay in the
original basis state is much larger than in the nearest ones. Comparing with
the result of the standard random matrix theory, one can say that there is a
noticeable difference (namely, the enhancement factor in Fig.3 is about 2.3,
instead of 3.0).

\section{The cascade model}

One of the important question is how the entropy of quantum isolated systems
increases in time at the transition to equilibrium. It is natural to define
the entropy of a many-body state through the Shannon entropy,
\begin{equation}
\label{S}S(t)=-\sum\limits_kw_k\ln w_k=-W_0\ln W_0-\sum\limits_{f\neq
0}w_f\ln w_f \,\,.
\end{equation}
Here $W_0(t)=\left| A_0(t)\right| ^2$ is the probability for the system to
be in the initial state, and $w_f\,(t)=\left| A_f\,(t)\right| ^2$ is the
probability to be in the basis state $\left| f\right\rangle $. In what
follows we assume that the initial conditions are $W_0(0)=1$ and $w_f(0)=0$,
therefore, the entropy is equal to zero for $t=0$.

In order to study the evolution of a many-particle system with two-body
interaction, it is convenient to introduce sub-classes for all basis states
in the following way. The {\it first class} contains those $N_1$ basis
states which are directly coupled with an initial state by the two-body
interaction given by matrix elements $H_{0f}\,$ of the interaction.
Correspondingly, the {\it second class} consists of $N_2$ basis states which
are coupled with the initial one in the second order of the perturbation,
this coupling is determined by $H_{0\alpha }H_{\alpha f}$ , etc.

Let us first consider the evolution for a large time $t\gg \frac {\Gamma}
{(\Delta E)^2}$ (below we assume the BW shape of the SF). For this case the
probabilities of the states in different classes can be determined by the
``probability conservation equations",
$$
\frac{dW_0}{dt}=-\Gamma W_0
$$
$$
\frac{dW_1}{dt}=\Gamma W_0-\Gamma W_1
$$
$$
.\,\,.\,\,.\,\,.\,\,.\,\,.\,\,.\,\,.\,\,.\,\,.\,\,.
$$
\begin{equation}
\label{Weq}\frac{dW_k}{dt}=\Gamma W_{k-1}-\Gamma W_k
\end{equation}
$$
.\,\,.\,\,.\,\,.\,\,.\,\,.\,\,.\,\,.\,\,.\,\,.\,\,.
$$
Here $W_k$ is the probability for the systems to be in the class $k$. The
first term $\Gamma W_{k-1}$ in the right-hand-side of (\ref{Weq}) is
responsible for the flux from the previous class, and the second term $%
\Gamma W_k$ describes the decay of the states in the class $k$, into the
next class $k+1$. We assume that the probability of the return to the
previous class can be neglected. This is a valid approximation if the number
of states $N_{k+1}$ in the next class is large in comparison with $N_k$ of
the previous class. This approach can be compared with those based on the
Caley tree model \cite{AGKL97} where the flow from each state goes into
other $M$ states, therefore, $N_k\simeq M^k$ with $M\gg 1$. Note that here
we consider a system which is far from the equilibrium. Indeed, if the
system is in the equilibrium, the probabilities of all states within the
energy shell defined by the relation $\left| E_f-E_0\right| \leq \Gamma $,
are of the same order, $w_f\simeq N_{pc}^{-1}$, with $N_{pc}$ as the total
number of states inside the energy shell. Therefore, in order to neglect the
return flux, one needs the condition $w_f = W_k/N_k\gg 1/N_{pc}$ to be
fulfilled.

Equations (\ref{Weq}) have the simple solution,
$$
W_0=\exp (-\Gamma t\,)
$$
\begin{equation}
\label{sol}W_n=\frac{(\Gamma t)^n}{n!}\exp \left( -\Gamma t\right) =\frac{
(\Gamma t)^n}{n!} W_0.
\end{equation}
The maximal probability $W_n=\frac{n^n}{n!}\exp (-n)\approx 1/\sqrt{2\pi n}$
to be in the class $n$ determined by the condition $\frac{dW_n}{dt}=0$ ,
occurs for $t=n/\Gamma $ , therefore, this solution (\ref{sol}) can be
considered as a {\it cascade} in the population of different classes .
Indeed, at small times $t\ll \tau \equiv 1/\Gamma $ the system is
practically in the initial state, at times $t\approx \tau \,$ the flow
spreads into the first class, for $t=n\tau $ it spreads into the $n-$th
class, etc. For an infinite chain one can easily check the normalization
condition,
\begin{equation}
\label{norm}\sum\limits_{n=0}^\infty W_n=\exp (-\Gamma
t)\,\sum\limits_{n=0}^\infty \frac{(\Gamma t)^n}{n!}=1.
\end{equation}

\section{Time dependence of the entropy}

The above expressions allow us to find the time-dependence of the entropy,
\begin{eqnarray}
\label{s(t)}
S(t)\approx -\sum \limits_{n=0}^\infty W_n \ln \frac{W_n}{N_n}=
\nonumber \\
\Gamma t \ln M + \Gamma t -e^{-\Gamma t} \sum\limits_{n=0}^\infty
\frac {(\Gamma t)^n}{n!} \ln \frac {(\Gamma t)^n}{n!},
\end{eqnarray}
where $w_f\approx \frac{W_n}{N_n}$ stands for the population of basis states
of the class $n$ with $N_n$ as the number of states in this class (in fact ,
for $t \sim n \tau$ one needs to count only the states inside the energy
shell since the population of the states outside the energy interval with $%
\left| E_f-E_0\right| > \Gamma $, is small, see Eq.(\ref{Wf})). Here we have
used the relations $N_n=M^n$ and $\sum\limits_{n=0}^\infty \frac{(\Gamma t)^n
}{n!}n= \Gamma t\exp (\Gamma t)$. Two last terms in the right-hand-side of
Eq.(\ref{s(t)} ) turn out to be smaller than the first one, therefore, one
can write,
\begin{equation}
\label{Sff} S(t)\approx \Gamma t \ln M [1 + f(t)]
\end{equation}
with some function $f(t) \ll 1$ which slowly depends on time.

In this estimate for the increase of entropy, we did not take into account
the influence of fluctuations of $w_f$. One can show that for gaussian
fluctuations of the coefficients $A_f$ with the variance given by their
mean-square values, for large number of principal components $N_{pc}(t)
\equiv \exp(S(t))$ the entropy should be corrected by a small factor of the
order of $\ln 2$ (see, for example, \cite{I90}).

If one neglects the second term in (\ref{Sff}), we obtain a linear increase
of the entropy, which means that the number of principal components $%
N_{pc}(t)$ increases exponentially fast with time. This behavior can be
compared with a linear increase of dynamical entropy $S_{cl}(t)$ in
classical chaotic systems where $S_{cl}(t)$ was found to be related to an
exponential divergence of close trajectories in the phase space ($%
S_{cl}(t)\propto \lambda t$ with $\lambda $ as the Lyapunov exponent, see,
for example \cite{Lyap}). The non-trivial point is that the linear increase
of entropy also occurs for systems without the classical limit, see recent
paper \cite{GPP00}.

Note, that for a small time the function $W_0(t)$ has the form $W_0(t)=\exp
(-\Delta_E^2\,t^2)$ (see (\ref{wgauss})), not the exponential dependence $%
\exp (-\Gamma t)$ . Therefore, one should modify the expression for the
entropy in order to make it valid for small times. For this, we replace $%
\Gamma t$ in Eq.(\ref{sol}) by a more accurate expression, $-\ln (W_0)$,
which gives $\Delta_E^2\,t^2$ for small time $t\ll \Gamma /\Delta_E^2$ and $%
\Gamma t\,$ for large time $t\gg \Gamma /\Delta_E^2$ , therefore,
\begin{equation}
\label{Wn}W_n=\frac{(\ln W_0^{-1})^n}{n!}W_0.
\end{equation}
It is easy to check that the normalization condition is fulfilled again,
\begin{equation}
\label{norm1}\sum_{n=1}^\infty W_n=\sum_{n=1}^\infty \frac{(\ln W_0^{-1})^n}{
n!}W_0=W_0\exp (\ln (W_0^{-1}))=1.
\end{equation}
For the entropy one obtains
\begin{equation}
\label{Snew}S=-\sum_{n=1}^\infty W_n\ln \left( \frac{W_n}{N_n}\right)
\end{equation}

At small time $t\ll \Gamma /\Delta_E^2\,$the entropy is given by two terms, $%
n=0$ and $n=1$ (direct transitions), therefore,
\begin{eqnarray}
\label{Sfinal}S(t)=-W_0(t)\ln W_0(t)-W_1(t)\ln \left( \frac{W_1}{N_1}\right)
\nonumber \\
\approx \Delta_E^2 t^2\,\left( 1+\ln \left( \frac{\Delta_E^2t^2}{N_1}
\right) ^{-1}\right)
\end{eqnarray}
This expression can be compared with the direct calculation based on the
relation $w_f=|H_{0f}t|^2$,
\begin{equation}
\label{pert}S(t)=\Delta_E^2 t^2\,+\,t^2\sum_f\left| H_{0f}\right| ^2\ln
\left( \frac 1{H_{0f}^2t^2}\right)
\end{equation}
There is an agreement between Eqs.(\ref{Sfinal}) and (\ref{pert}) since $%
\Delta_E^2=\sum_f\left| H_{0f}\right| ^2=N_1\,\overline{H_{0f}^2}$ .

Let us now discuss the whole time dependence of the entropy, including large
times $t$ when the system is close to the equilibrium. In a finite system of
particles any basis state can be reached, starting from an initial state, in
several ``interaction steps'' $(H_{0\alpha}H_{\alpha \beta }H_{\beta \gamma
}\,.\,.\,.)$ . For example, in the system of $n=6$ particles three steps is
needed since the two-body interaction can not move more than two particles
from one basis state to another. If the number of classes $n_c$ is finite,
the states in the last class do not decay (there is no term $-\Gamma W_{n_c}$
in the last equation in (\ref{Weq})), and the probability to be in the last
class is determined from the normalization condition $W_{n_c}=1-%
\sum_{k=0}^{n_c -1}W_k$. The additional condition is that the considered
basis states should be inside the energy interval, $\left| E_f-E_0\right|
\leq min(\Gamma , \Delta E)\,$, thus limiting the number $N_n$ of the basis
states in each class. Note that the value of $W_n$ is restricted from below
by the equilibrium relation $W_n\geq N_n/N_{pc}$ . These limitations make
``exact'' expression for the entropy very complicated. Instead, we can
propose the following simple expression which is approximately valid in
systems with a small number of classes ($n_c \sim 1$),
\begin{equation}
\label{Sappr}S(t)=-W_0(t)\ln W_0(t)\,-(1-W_0(t))\ln \left( \frac{(1-W_0(t))}{%
N_{pc}}\right)
\end{equation}
This expression takes into account the normalization condition $%
\sum\limits_{f\neq 0}w_f = 1-W_0$ and has a reasonable behavior for both
small and large times.

\section{Numerical results for the entropy}

Now we compare the obtained analytical expressions with numerical data for
the TBRI model. For the case when the strength function (SF) has the
Breit-Wigner form, the time dependence of the entropy is shown in Fig.4 for
the parameters of Fig.2, $n=6,\,m=12,\,V_0^2 \approx 0.003, \Gamma_0 \approx
0.50,\,\Delta_E \approx 1.16$, with the average over $N_g=2$ Hamiltonian
matrices.

\begin{figure}[htb]
\vspace{-1.cm}
\begin{center}
\hspace{-2.6cm}
\epsfig{file=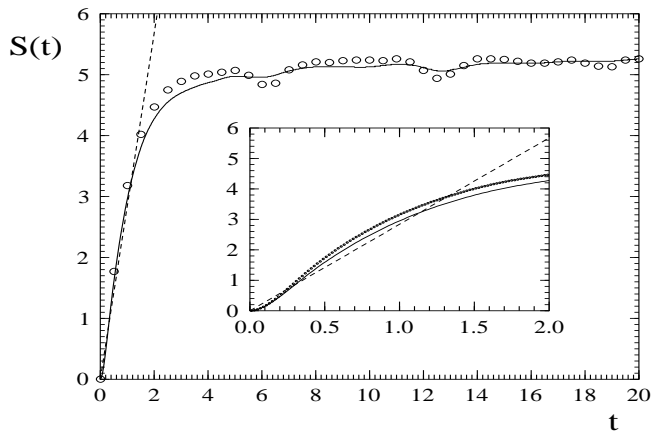,width=3.0in,height=3.3in,angle=-90}
\vspace{-0.5cm}
\caption{
Entropy versus time for the TBRI model in the case when the SF has the
standard Breit-Wigner form. The parameters are the same as in Fig.2, with $%
N_g=2$. The circles stand for numerical data, the solid curve is
the analytical expression (\ref{Sappr}), and the dashed line
represents a linear slope according to an approximate expression
(\ref{SBW}). In the inset the same is shown for a smaller time
scale. }
\end{center}
\end{figure}
The number $M$ of the basis states directly coupled by the random two-body
interaction, is determined by the expression \cite{FGI96},
\begin{equation}
\label{M}M=n(m-n)+\frac{n(n-1)(m-n)(m-n-1)}4,
\end{equation}
where the first term gives the number of one-particle transitions, and the
second stands for two-particles transitions. In our case of $n=6$ particles
and $m=12$ orbitals, the total number of basis states is $N=924$ and $M=261$%
. The {\it effective number} of classes in the cascade model can be
determined from the relation $M^{n_c}=N$. This gives $n_c=\ln N/\ln M\approx
1.2$. Thus, we can use the simple expression (\ref{Sappr}) to describe the
dependence of the entropy on time analytically. The data in Figs.4-5
demonstrate an excellent agreement between the numerical and analytical
results.

We should note that the theoretical dependence ({\ref{Sappr}) which gives
quite good approximate description of the data on the whole time scale, has
the parameter $N_{pc}$ (effective number of principal components in the
stationary distribution $w_f(t\rightarrow \infty))$ which is related to the
limiting value of the entropy, $N_{pc}=\ln S(\infty)$. It can be estimated
analytically as discussed above from the width of the energy shell; when
plotting the solid curves in Figs.4-5 we have used the exact value found
numerically.}

To avoid confusion, we should explain that the {\it actual number} of
classes in the case of $n=6$ particles and $m=12$ orbitals is equal to $3$
since all basis states can be populated in the third order in the two-body
interaction. However, the number of states in the second, $k=2$, and third, $%
k=3$, classes are much smaller than it may follow from the exponential
relation $N_k=M^k$ (in practice, this relation may be justified for a large
number of particles only). This is the reason why the one-class formula (\ref
{Sappr}) works so well.

It is also instructive to compare the entropy with the linear time
dependence,
\begin{equation}
\label{SBW}S(t)=\Gamma \,t\,\ln M
\end{equation}
that stems from Eq.(\ref{Sff}) if the first term is taken only. This
dependence globally corresponds to the data on some time scale, however, the
actual dependence for $S(t)$ clearly differs from the linear one (see inset
in Fig.4). Note that the quadratic increase of energy occurs on a very small
time scale only. As for the oscillations of the entropy for a very large
time close to an equilibrium, this phenomenon will be discussed below.

\begin{figure}[htb]
\vspace{-3.8cm}
\begin{center}
\hspace{-2.5cm}
\epsfig{file=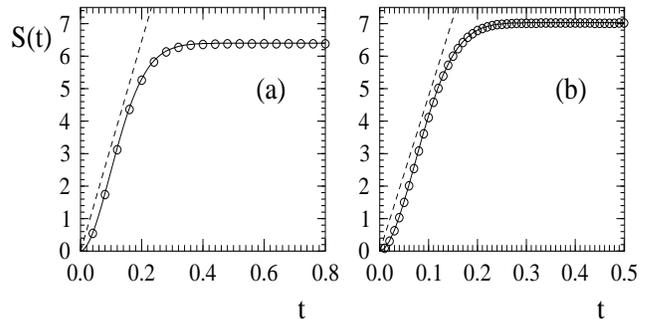,width=4.0in,height=3.0in,angle=-90}
\vspace{-2.9cm}
\caption{Time dependence of the entropy for the TBRI model
when the strength function is of the Gaussian form: (a) $n=6,
m=12, V_0\approx 0.083, \Gamma_0\approx 10.5, \Delta_E\approx
5.8$, and (b) $n=7, m=13, V_0\approx 0.12, \Gamma_0\approx 14.6,
\Delta_E \approx 8.13$. Circles are numerical data for $N_g=2$, solid
curves stand for the approximate expression (\ref {Sappr}), and
dashed lines represent the linear dependence (\ref{SG}).}
\end{center}
\end{figure}

For a strong interaction, when the form of the SF is very close to the
Gaussian, numerical data are reported in Fig.5 for $n=6, m=12$ and for $n=7,
m=13$. The interaction strength is chosen in order to have the same ratio $%
\Gamma_0/\Delta_E \approx 1.8$, as in Fig.4.

In this case the half-width of the strength function is determined by $%
\Delta_E$ since $\Gamma_0=2\pi \rho_f H_{0f}^2$ is larger than $\Delta_E$.
As a result, the role of $\Gamma$ in the expressions (\ref{Sff}) and (\ref
{Sappr}) plays the width $\Delta_E$. In both cases numerical data give
strong evidence of a linear increase of the entropy,
\begin{equation}
\label{SG} S(t)=\Delta_E \, t \, \ln M,
\end{equation}
before the saturation. It is clearly seen that this analytical estimate
gives a correct value for the slope of $S(t)$. The shift is due to the
initial time scale where the time dependence is quadratic, this fact is
neglected in the estimate.

It should be pointed out that the linear dependence of $S(t)$ in Fig.5 is
much more pronounced than in the BW-region (compare with Fig.4 for which the
SF is of the Breit-Wigner form). Our results indicate a clear difference
between the two cases related to the Breit-Wigner and Gaussian forms of the
SF. This point is somehow supported by recent studies \cite{CIK00,doron}
where it was shown that for a relatively weak interaction ($\Gamma _0$ is
small in comparison with $\Delta _E$) resulting in the Breit-Wigner form of
the SF, there is no detailed quantum-classical correspondence for the
evolution of wave packets in the energy space. On the other hand, in the
Gaussian region (with the Gaussian form for the SF), the detailed
quantum-classical correspondence is possible \cite{CIK00,doron}. In the
latter case one can expect a linear growth for the entropy, as it was found
in classical models \cite{Lyap}. The principal difference between these two
cases (in what concerns the quantum-classical correspondence and the
possibility of the localization in the energy shell) for the first time has
been discussed in Ref.\cite{CCGI96}.

\section{Width of packets and inverse participation ratio}

The width of the wave function in the basis representation can be measured
through the variance,
$$
\Delta ^2(t)=\sum_f\left( n_f-n_0\right) ^2\left| A_f\,(t)\right|^2
=\left(1-W_0(t)\right) \left| \Delta _f\,(t)\right| ^2;
$$
\begin{equation}
\label{Delta}\left| \Delta _f\,(t)\right|^2= \frac{\sum_f\,(n_0-n_f)^2\left|
A_f\,(t)\right| ^2}{\sum_f\,\left| A_f\,(t)\right| ^2},
\end{equation}
where $n_f$ and $n_0$ label corresponding basis states, $f\neq 0$, and we
have used the normalization condition $\sum_f\left| A_f\,(t)\right|
^2=1-W_0(t)$ . The function $\left| \Delta _f\,(t)\right| ^2\,$ is a slow
function of time, it changes from the effective band-width of the
Hamiltonian matrix, entirely determined by the matrix elements $H_{0f}$, to
the final width of the wave packet in the basis representation, which is
defined by the width of the energy shell (approximately equal to $\sqrt{2}%
\Delta E$). Therefore, the leading time dependence is given by the term $%
1-W_0(t)$.

For relatively small times before the saturation, we can use the simple
estimate $w_f\approx |H_{0f}t|^2$, which results in the following quadratic
dependence,
\begin{equation}
\label{Deltat}\Delta ^2(t)\approx t^2\sum_f\left( n_f-n_0\right)
^2H_{0f}^2=t^2V_0^2\Delta _0^2.
\end{equation}
Here $\Delta _0$ is some constant related to an effective band-width of the
Hamiltonian matrix. The linear dependence for the width of packet, $\Delta
(t)=t\,V_0\,\Delta _0$, corresponds to a generic ballistic-like behavior of
wave packets found for the WBRM model \cite{CIK00}.

Note that the band-width of the Hamiltonian matrix can be much larger than
the final width of the wave packet due to the dependence of the latter on
the interaction strength. Therefore, the linear increase of $\Delta (t)$ can
be very fast, and quickly it becomes saturated on the time scale of the
applicability of the expansion in $Ht$, see Eq. (\ref{Deltat}).

Before comparing the obtained expressions with numerical data, let us first
analyze the time dependence of the number of principal components $N_{pc}(t)$
for wave packets in the basis representation. It is natural to define $N_{pc}
$ through the entropy, $N_{pc}(t)=\exp (S(t))$. This definition has been
widely used in different applications, see for example, \cite{I90}.

The number of principal components can be also defined through the inverse
participation ratio $l_{ipr}$:
\begin{equation}
\label{NPC} \left( l_{ipr}\right) ^{-1} =\sum \left| A_f\right| ^4\approx
\sum_k \frac{W_k^2}{N_k}\approx W_0^2\sum_k\frac{(\ln W_0^{-1})^{2k}}{k!k!N_k%
}.
\end{equation}
Here we used Eq.(\ref{Wn}) for $W_k$. The sum in Eq.(\ref{NPC}) gives the
following result for the infinite number of classes (this may be a
reasonable approximation for time $t<<n_c\tau $)
\begin{equation}
\label{NPC1}\left( l_{ipr}(t)\right) ^{-1}=W_0^2\,I_0\left( \frac{2\ln
(W_0^{-1})}{\sqrt{M}}\right),
\end{equation}
where $I_0(z)$ stands for the modified Bessel function and we used the
relation $N_n=M^n$.

This expression has the following asymptotics,
\begin{eqnarray}
\label{NPC2}\left( l_{ipr}(t)\right) ^{-1}=W_0^2\left( 1+\frac 1{N_1}(\ln
W_0^{-1})^2\right)
\nonumber \\
 \approx 1-2t^2(\Delta E)^2(1+\frac 1{N_1}),
\end{eqnarray}
for small time, and
\begin{equation}
\label{NPC3}\left( l_{ipr}(t)\right) ^{-1}=\exp \left( -2\Gamma (1-\frac 1{
\sqrt{M}})\,t\right)
\end{equation}
for large time. Therefore, $N_{pc}$ defined through the inverse
participation ratio $l_{ipr}$ also may have an interval of exponential
increase in time (if the number of classes $n_c$ is not small and we can
extend the summation over $k$ in Eq. (\ref{NPC}) to infinity) . Here we
again neglected the fluctuations of $A_f(t)$ which may increase the value of
$l_{ipr}^{-1}$ up to the factor $3$.

For a system with small number of classes on can suggest the following
approximate expression,
\begin{equation}
\label{Nfinal}\left( l_{ipr}(t)\right) ^{-1}=W_0^2+\frac{(1-W_0)^2}{%
l_{ipr}(\infty )}.
\end{equation}
This expression takes into account the normalization condition $%
\sum\limits_{f\neq 0}w_f=1-W_0$ and has a reasonable behavior for both small
and large time.

In any system with finite number of particles the energy shell
contains finite number of basis states. For a stationary chaotic
state the number of principal components is estimated as
$N_{pc}^{st}\sim \Gamma /D\,$ where $D$ is the mean energy
distance between many-particle levels (we assume here that the
spreading width $\Gamma \,$ of exact eigenstates is less than
$\Delta_E$). In the non-stationary problem this leads to the
saturation of $ N_{pc}(t)$ to the value $N_{pc}(\infty )\simeq 2
N_{pc}^{st}$, and to the maximal value $S\approx \ln
\,N_{pc}(\infty )$ for the entropy (see above and in Ref.
\cite{F2000}).

Numerical data for the TBRI model for the case when the SF is of
the Breit-Wigner form are summarized in Fig.6 for the same
parameters as in Fig.2 and Fig.4. Three quantities are plotted
here: the width $\Delta (t)$, the number of principal components
$N_{pc}(t)=\exp(S(t))$ and $l_{ipr}(t) =(\sum|A_f(t)|^4)^{-1}$
determined by the inverse participation ratio. For $N_{pc}(t)$ two
curves are given, one is due to the analytical expression
(\ref{Sappr}) for the entropy, and another is computed directly
from the evolution of the TBRI model, with additional average over
$N_g=2$ number of realization of random Hamiltonian.

From the reported data one can see that the time dependence of the
width $\Delta (t)$ of packets is quite simple. Namely, on the
first very short time scale the increase of the width is linear in
time (see also inset in Fig.6 where $\Delta (t)$ is shown on this
time scale), and after, the width quickly saturates. This behavior
is in the correspondence with analytical estimates discussed
above, and with numerical results found in the WBRM model
\cite{CIK00}.

In contrast to this time dependence, the increase of the number of
principal components $N_{pc}(t)$ is very different, both for
$N_{pc}(t)$ defined by the entropy and for the $l_{ipr}(t)$
determined by the inverse participation ratio. Indeed, both
$N_{pc}(t)$ and $l_{ipr}(t)$ increase slowly in time before the
saturation to their limit values. The absolute difference between
these two quantities is not important since the definition of
$l_{ipr}$ is given up to some factor which is sensitive to the
type of fluctuations in $ A_f(t)$. As we already noted, the
gaussian fluctuations decrease the value of $l_{ipr}$ by the
factor $3$.

\begin{figure}[htb]
\vspace{-1.5cm}
\begin{center}
\hspace{-3.2cm}
\epsfig{file=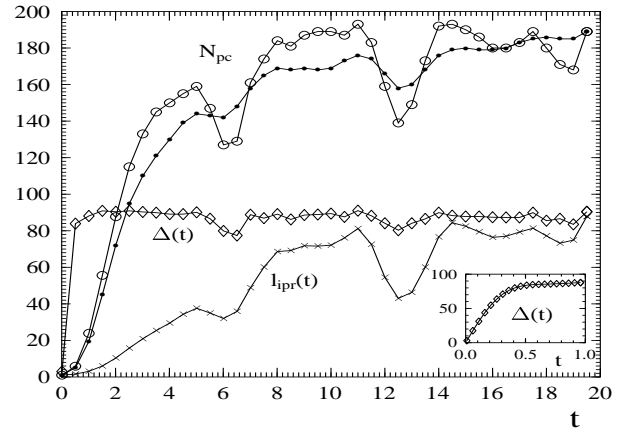,width=3.0in,height=3.3in,angle=-90}
\vspace{-0.8cm}
\caption{
Time dependence of different quantities for the TBRI model in the
case of the BW-form of strength functions. The parameters are the
same as in Fig.4, with $N_g=2$. Circles stand for numerical data
for $N_{pc}=\exp(S(t))$ with $ S(t)$ taken from Fig.4, dots
correspond to the analytical expression for $ N_{pc}(t)$ with
$S(t)$ from Eq.(\ref{Sappr}), triangles represent numerical data
for the width $\Delta (t)$, and squares are numerical results for
$ l_{ipr}(t)$. The width $\Delta (t)$ on a smaller time scale is
shown in the inset. }
\end{center}
\end{figure}

One of the most interesting facts which can be drawn from these data is a
big difference for characteristic time scales which correspond to the
saturation. Indeed, if the width $\Delta (t)$ completely saturates at time $%
t \approx 0.5$, both $N_{pc}$ and $l_{ipr}$ manifest a very slow saturation
by the time $t \approx 20$. This means that the mechanism for the width
increase is different from that responsible for the increase of the number
of principal components in the wave packet.

To explain this phenomenon let us consider the initial time scale for the
time dependence of $N_{pc}$ and $l_{ipr}$, where one can detect an
approximate linear increase of the entropy $S(t)$. The very point is that at
small times $t\leq \tau $ the wave function has a large number of holes
since only directly connected basis states are populated. For a system with
a large number $n$ of particles the fraction of these states is
exponentially small $(\sim \exp (-n))$ due a two-body nature of the
interaction. With an increase of time, for $t\geq\tau =\Gamma_0 ^{-1}$ , the
states in other classes start to be filled and the holes begin disappear.
This stage for $t< n_c\tau$ corresponds to an exponential increase of the
number of principal components and linear increase of the entropy.

It is instructive to analyze the evolution of wave packets on a smaller time
scale of the ballistic spread, see Fig.7. These data confirm theoretical
expectations according to which at small times only those basis states that
belong to the first class, are involved in the dynamics. Indeed, large gaps
are clearly seen in the distribution of $w_f$, which persist during all the
time of the ballistic-like spread. These gaps reduce the number of the
principal components, however, they are not important for the calculation of
$\Delta(t)$, see Eq.(\ref{Delta}).

An important peculiarity of the wave dynamics is that initially {\it all}
basis states of the first class are excited. One can see that for a very
small time $t=0.2$ the whole available region $1<f<N$ is filled with
approximately the same amplitudes $w_f=|H_{0f} t|^2 $. With an increase of
time, the amplitudes grow and form the envelope of the packet, in accordance
with Eq.(\ref{Wf}). One should stress that the quadratic time dependence for
the second moment of a packet on this short-time scale occurs not due to a
linear spread of the front of the wave packet, but due to a specific growth
of the amplitudes of those basis states which are located inside the energy
shell.

\begin{figure}[htb]
\vspace{-1.2cm}
\begin{center}
\hspace{-2.3cm}
\epsfig{file=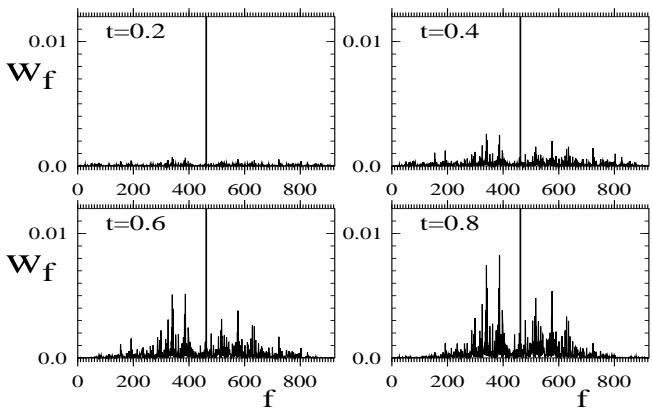,width=2.8in,height=3.in,angle=-90}
\vspace{-0.8cm}
\caption{
Wave packet $w_f(t)$ for the TBRI model at different times
$t=0.2,\,0.4,\,0.6.\,0.8$ for the parameters of Fig.7. One
particular Hamiltonian matrix is used without any additional
average. }
\end{center}
\end{figure}

The remarkable effect is a kind of oscillations for all quantities of Fig.6.
Similar oscillations (however, not so strong) are also present in the time
dependence of the entropy $S(t)$, see Fig.4. Since the period $T\approx 6.5$
of these oscillations is much larger than the time scale $t\approx 0.5$ of
the ballistic spread of wave packets, it is clear that this effect is
entirely related to the dynamics in the Fock space formed by different
classes. The origin of these oscillations can be explained in terms of the
cascade model discussed in Section III. Indeed, one can expect a strong
effect of a {\it reflection} due to the finiteness of the Fock space. The
first reflection occurs for $t_0 \approx n_c (\Gamma_0)^{-1}$, therefore,
the period of oscillation is $T \approx 2 t_0$. One can see that this
estimate gives the correct result for $T$ with $n_c\approx 1.5$. The latter
value is close to our rough estimate $n_c\sim 1.2$ for an effective number
of classes.

To compare the data for $l_{ipr}$ defined by the inverse participation ratio
with the analytical expression (\ref{Nfinal}), we have plotted separately
both results for a larger time scale $t\leq 40$ in Fig.8. One can see that
our estimate (\ref{Nfinal}) gives quite accurate description of the data on
a large time scale up to $t\approx 20$. After this time, the saturation
occurs and all local time dependence may be treated as fluctuations around
the limiting value. The data presented in this figure give strong evidence
of the effectiveness of our analytical approach.

\begin{figure}[htb]
\vspace{-1.3cm}
\begin{center}
\hspace{-2.3cm}
\epsfig{file=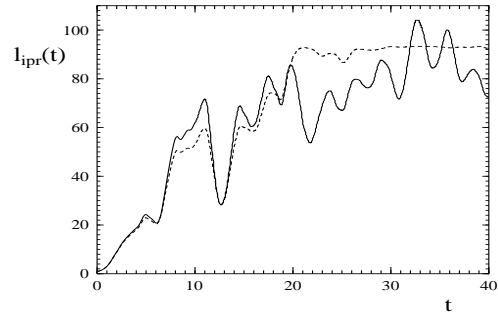,width=2.2in,height=2.4in,angle=-90}
\vspace{-0.8cm}
\caption{
Time dependence of $l_{ipr}(t)$ for the parameters of Fig.6 on a
larger time scale for one Hamiltonian matrix. Solid curve stand
for numerical data, and the dashed curve corresponds to the
analytical expression (\ref{Nfinal}). }
\end{center}
\end{figure}

Let us now come to the case when the form of the strength function
is close to the Gaussian. In Fig.9 one can see that the analytical
expressions connecting the probability $W_0(t)$ to stay in an
initial basis state, with the time dependence of the number of
principal components $N_{pc}(t)=\exp (S(t))$ and $l_{ipr}(t)$,
give correct global description of numerical data. When comparing
with the previous case of the BW-form of the SF, see Figs.6,8, one
can conclude that for a strong interaction (when the Gaussian
dependence for the SF emerges) the behavior of all quantities do
not reveal strong oscillations. This effect is related to the fact
that the time scale for the saturation of the width of packets is
of the same order as for the saturation of both $N_{pc}$ and
$l_{ipr}$. In such a case the effect of reflection in the Fock
space is suppressed by a strong spread of packets in the energy
shell.

\begin{figure}[htb]
\vspace{-1.2cm}
\begin{center}
\hspace{-2.3cm}
\epsfig{file=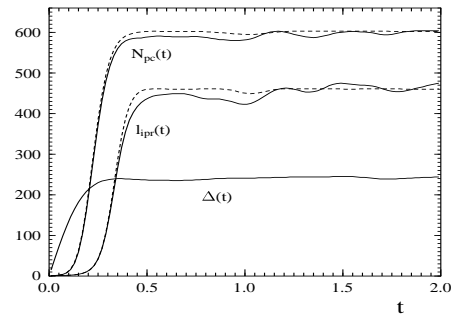,width=2.2in,height=2.4in,angle=-90}
\vspace{-0.5cm}
\caption{
The same quantities as in Fig.6 for a strong interaction with the
Gaussian form of the SF. The parameters for the TBRI model are the
same as in Fig.5a: $n=6, m=12, V_0\approx 0.083, \Gamma_0\approx
10.5, \Delta_E \approx 5.8$. Solid curves correspond to numerical
data, dashed curve stand for the corresponding analytical
expressions: Eq.(\ref{Nfinal}) for $l_{ipr}(t)$, and
Eq.(\ref{Sappr}) for $S(t)$ in the definition
$N_{pc}(t)=\exp(S(t))$. }
\end{center}
\end{figure}

One can see, that there is a quite strong difference in the time-dependence
of all discussed above quantities for the two extreme cases of the
Breit-Wigner and Gaussian forms of the strength function. In the BW-case the
two effects (ballistic-like spread of packets and the cascade-like evolution
in the Fock space) have very different time scales, and both these effects
can be distinguished in the dynamics. Contrary, in the second case of the
Gaussian form of the SF, two time scales are comparable. Therefore, the two
effects coexist on the same time scale and, as a result, the global
time-dependence turns out to be much simpler.

\section{Comparison with Wigner Band Random matrices}

In this Section we discuss numerical results obtained for the WBRM model for
the same quantities as the considered above. The dynamics of wave packets in
this model has been recently studied in \cite{CIK00} in the connection with
the problem of quantum-classical correspondence. Here, instead, we
concentrate our attention on the correspondence between the evolution of
packets in the WBRM and TBRI models.

As was mentioned, the WBRM model is quite close to the TBRI model. It
consists of two parts, one of which is a diagonal matrix with increasing
``energies" $\epsilon_j$ and another is a band matrix $V_{ij}$,
\begin{equation}
\label{wbrm}H_{ij}=\epsilon _j\delta _{ij}+V_{ij},
\end{equation}
where $\delta_{ij}$ is the delta-function. In original papers \cite{W55} the
``unperturbed spectrum'' was taken in the form of the ``picked fence'', $%
\epsilon _j=jD$, where $D=\rho_0^{-1}$ is the spacing between two close
energies and $j$ is a running integer number. We consider here the case with
random values $\epsilon _j$ with the mean spacing $D$, reordered in an
increasing way, $\epsilon_{j+1} > \epsilon_j$. As for the off-diagonal
matrix elements $V_{ij}$, they are assumed to be gaussian random and
independent variables inside the band $|i-j|\le b$, with the zero mean, $%
<V_{ij}=0>$, and the variance $<V_{ij}^2>=V_0^2$. Outside the band, the
matrix elements vanish. Thus, the control parameters of this model are the
ratio $V_0/D$ of a typical matrix element to the mean level spacing, and the
band-width $b$. As one can see, the first term in (\ref{wbrm}) corresponds
to a ``mean field'' $H_0$, and the interaction $V$ has a finite energy range.

For the first time the SF for the WBRM has been analyzed in Ref. \cite{W55}.
It was analytically found that the form of the strength function essentially
depends on one parameter $q=\frac{\rho _0^2V_0^2}b$ only. Wigner has proved
\cite{W55} that for a relatively strong perturbation, $V_0\gg D$, in the
limit $q\ll 1$ the form of the LDOS is the Lorentzian,

\begin{equation}
\label{BW} W_{BW}\,(\tilde E)= \frac 1{2\pi }\,\frac{\Gamma _{BW}}{\tilde
E^2+\frac {1}{4} \Gamma _{BW}^2},\,\,\,\,\,\,\,\,\tilde E=E-D\,j,
\end{equation}
which nowadays is known as the Breit-Wigner (BW) dependence. Here the energy
$\tilde E$ refers to the center of the distribution. The half-width $%
\Gamma_{BW}$ of the distribution (\ref{BW}) is given by the Fermi golden
rule,
\begin{equation}
\label{BWgam}\Gamma _{BW}=2\pi \rho _0V_0^2
\end{equation}

In other limit $q\gg 1\,$ the influence of the unperturbed part $H_0\,$ can
be neglected and the shape of the SF tends to the shape of density of the
states of the band random matrix $V$, which is known to be the semicircle.

Recently, Wigner's results have been extended in \cite{FCIC96} to matrices $%
H $ with a general form of $V$, when the variance of the off-diagonal matrix
elements decreases smoothly with the distance $r=\left| i-j\right| $ from
the principal diagonal. In this case the effective band size $b\,$ is
defined by the second moment of the envelope function $f(r)$ . Another
important generalization of the WBRM studied in \cite{FCIC96}, is an
additional sparsity of the matrix $V$, which may mimic realistic
Hamiltonians. In such a form, the WBRM model is closer to the TBRI model,
however, in the latter the sparsity of the interaction is due to a two-body
nature of the interaction. As a result, the positions of zero elements are
not completely random as in the WBRM model, see details in \cite{FGI96,I00v}.

Random matrix models of the type (\ref{wbrm}) are very useful for the
understanding generic properties of the SF. The condition for the SF to be
of the BW form in the WBRM model has simple form \cite{FCIC96},
\begin{equation}
\label{range1}D \ll \,\Gamma _{BW}\ll \Delta_b;\,\,\,\,\,\, \Delta_b=b D.
\end{equation}
The left part of this relation indicates the non-perturbative character of
an interaction, according to which many of unperturbed basis states are
strongly coupled by an interaction. On the other hand, the interaction
should not be very strong, namely, the width $\Gamma_{BW}$ determined by Eq.(%
\ref{BWgam}), has to be less than the width $\Delta_b$ of the interaction in
the energy representation. The latter condition is generic for systems with
finite range of the interaction $V$. One should stress that, strictly
speaking, the BW form (\ref{BW}) is not correct since its second moment
diverges. As was shown in \cite{W55,FGGK94}, outside the energy range $%
|\tilde E|> \Delta_b$ the SF in the model (\ref{wbrm}) decreases with the
energy faster than a pure exponent.

Note, that in the TBRI model the energy scale $\Delta _b$ is irrelevant
since there is no sharp border of the interaction and $\Delta _b$ is of the
order of the whole energy spectrum. For this reason, instead of $\Delta _b$
it is more convenient to use the variance $\Delta _E^2$ of the SF which may
have the classical limit \cite{CCGI96,CIK00}. The latter quantity can be
expressed through the off-diagonal matrix elements of the interaction, $%
\Delta _E^2=\sum_jV_{ij}^2$ for $i\neq j$, therefore, $\Delta _E^2=2bV_0^2$.
As a result, we have $\Delta _b=\pi \Delta _E^2/\Gamma _{BW}$ and Eq.(\ref
{range1} ) can be written as
\begin{equation}
\label{range2}D\ll \,\Gamma _{BW}\ll \Delta _E\sqrt{\pi }.
\end{equation}

Numerical data \cite{CFI00} for the WBRM model show that on the
border $\Gamma _{BW}\approx 2\Delta _E$ the form of the SF is
quite close to the Gaussian, and this transition from the BW
dependence to the Gaussian-like turns out to be quite sharp.
Although the extreme limit of a very strong interaction, $q\gg 1,$
(or, the same, $\Gamma _{BW}\gg \Delta _E$) has been studied by
Wigner in the WBRM model (\ref{wbrm}) , the semicircle form of the
SF seems to be unphysical. Indeed, this form is originated from
the semicircle dependence of the total density $\rho _V(E)$
defined by $V$ only, therefore, when neglecting the term $H_0$.
Thus, the case $V\gg H_0$ in terms of the TBRI model means that
the residual interaction is much stronger than the mean-field part
$H_0$, which is physically irrelevant.

\begin{figure}[htb]
\vspace{-1.cm}
\begin{center}
\hspace{-3.2cm}
\epsfig{file=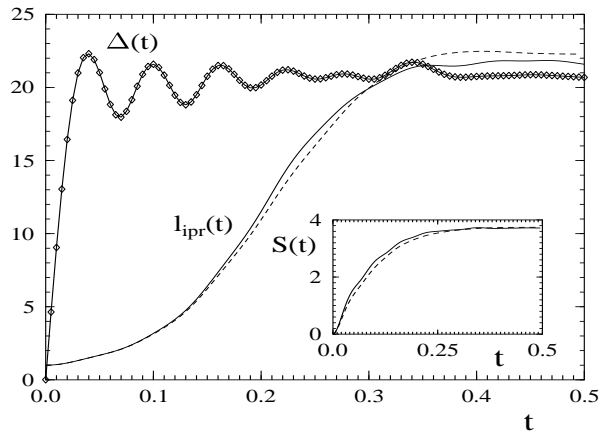,width=3.0in,height=3.3in,angle=-90}
\vspace{-0.5cm}
\caption{
Time dependence of the entropy $S(t)$ (in the inset), the width of
the packet $\Delta(t)$ and the number of principal components
$l_{ipr}(t)$ for the WBRM model for the case of the BW-form of the
SF. The parameters are: $ N=924, b=110, D=1.0, V_0=1.0$,
correspondingly, $\Gamma_{BW}\approx 6.28,
\Delta_E \approx 14.8$. Solid curves for $S(t)$ and $l_{ipr}(t)$ are
numerical data, dashed curves represent analytical expressions
(\ref{Sappr}) (for $S(t)$) and (\ref{Nfinal}) (for $l_{ipr}$).
Diamonds, connected by the solid line, correspond to numerical
data for $\Delta(t)$. }
\end{center}
\end{figure}

Numerical data for the WBRM model in the case of the BW-dependence
of the SF are given in Fig.10. When comparing with the
corresponding quantities discussed above for the TBRI model (see
Fig.6) we should note the following. First of all, one can see
that the simple analytical expression (\ref{Sappr} ) gives a
correct description of the increase and saturation of the entropy
$ S(t)$. The same we can say about the expression (\ref{Nfinal})
for the number of principal components defined through the inverse
participation ratio.

Second, we would like to stress that the global time dependence for all
quantities is quite similar to that found for the one-class variant of the
TBRI model. Relatively simple structure of the Wigner band random matrices
allows one to perform a detailed comparison of the data with analytical
estimates. Indeed, the application of the relation (\ref{Deltat}) for the
WBRM model gives,
\begin{equation}
\label{WBRMwidth}\Delta ^2(t)=\frac 23t^2V_0^2b^3,
\end{equation}
see also Ref.\cite{IKPRT97}. Therefore, for the parameters of Fig.10 we have
$\Delta (t)=Bt$ with $B=\sqrt{\frac 23}b^{3/2}V_0\approx 950$ which is in a
good agreement with numerical data. We can also find the critical time $t_d$
after which the ballistic spread of the packet terminates. For this, we
estimate the maximal width $\Delta _m$ of packet via the width $\Delta _E$
of the SF, $\Delta _m\approx \sqrt{2}\Delta _E$. This leads to the estimate $%
t_m\approx \sqrt{6}/b$, therefore, for Fig.3 we have $t_m\approx 21$ which
perfectly corresponds to the data. The latter estimates have been also
checked for other values of $V_0$ and $b$, with the same good correspondence
between simple estimates and numerical data.

Comparing global time dependence of the quantities presented in Fig.10 with
the results for the TBRI model, see Fig.6, one can see that the main
difference is the type of oscillations for the width of packets $\Delta (t)$
. Namely, in contrast to the TBRI model where the period of oscillations is
much larger than the time scale $t_m$ of the ballistic spread, in Fig.10 the
period $T$ is just defined by the ballistic spread, $T\approx 2t_m$. This
very fact demonstrate the principal difference between the two models.

Indeed, for the WBRM model there is no specific evolution in the Fock space
which is due to a two-body nature of interaction. Formally, the cascade
model can be applied to the WBRM model with the number of classes $n_c=1$,
since all states within the energy band $\Delta _b=bD$ start to be involved
in the dynamics immediately. This means that in contrast to the TBRI model,
in the WBRM model there is only one mechanism for the oscillations, namely,
the reflection inside the energy shell which is populated {\it ergodically},
when time is running. No oscillations are detected for the number of
principal components (the data for larger times are not shown), this
confirms our conclusion about one kind of the reflection from the edges of
the energy shell. It is interesting to note that the number of principal
components does not reveal noticeable oscillations on the time scale of the
ballistic spread since on this scale the value of $N_{pc}$ is very small.

The difference between these two models can be also seen when comparing the
structure of wave packets in the basis representation at some instant times
before the saturation (compare Fig.11 with Fig.7). In contrast to Fig.7
where many ''holes'' can be realized in the distribution $w_f$, for the WBRM
model the filling of the available energy range of size $2bD$ occurs
ergodically. In both cases very strong fluctuations are present, which are
expected to be gaussian, see discussion in \cite{FI97}. It is important to
stress, that in order to reveal this difference, we should avoid the
ensemble average which washes out the presence of holes (if different
Hamiltonian matrices have different unperturbed spectrum). This fact
reflects one of basic peculiarities of the TBRI model, namely, the
non-ergodic character of the matrices (the average over the spectrum inside
one (very big) matrix may give completely different result from that
obtained by an ensemble average, see references in the review \cite{brody}).

\begin{figure}[htb]
\vspace{-1.cm}
\begin{center}
\hspace{-2.3cm}
\epsfig{file=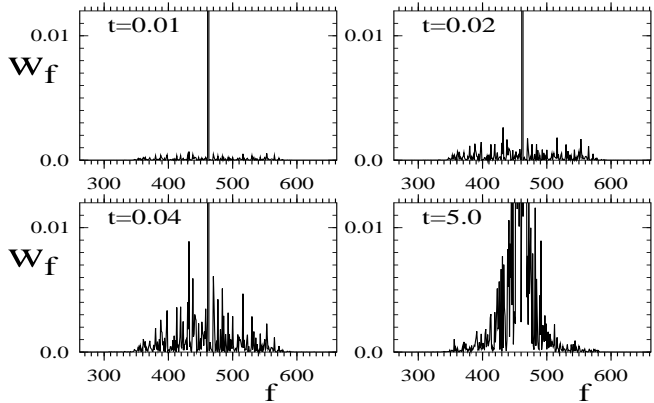,width=2.8in,height=3.in,angle=-90}
\vspace{-0.5cm}
\caption{
Wave packet $W_n(t)$ for the WBRM model at different times
$t=0.01,\,0.02,\,0.04,\,5.0$ for the parameters of Fig.10. One
particular band random matrix is used without any additional
average. }
\end{center}
\end{figure}

\begin{figure}[htb]
\vspace{-1.cm}
\begin{center}
\hspace{-3.2cm}
\epsfig{file=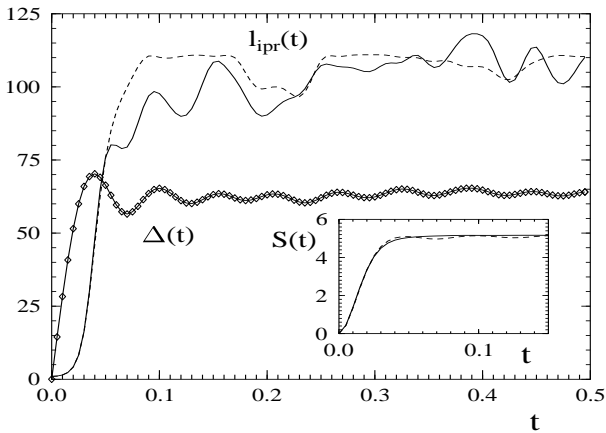,width=3.0in,height=3.3in,angle=-90}
\vspace{-0.5cm}
\caption{
The same as in Fig.10 for the case when the form of the SF is
close to the Gaussian. The parameters are: $N=924, b=110, D=1.0,
V_0=3.0$, correspondingly, $\Gamma_{BW}\approx 56.5, \Delta_E
\approx 45.5$. }
\end{center}
\end{figure}
Finally, we present the data for the WBRM in the case of the
Gaussian form of the strength function, see Fig.12. Here we also
can see oscillations for the width of packets. As for the number
of principal components found from the inverse participation
ratio, numerical data manifest the oscillations, as well, with the
same period as for the width of packets. In average, the numerical
data for $S(t)$ and $l_{ipr}$ are well described by the simple
analytical expressions relating these quantities to the
probability $W_0(t)$ to stay in the initial state.

The most interesting result which can be drawn from the numerical data
presented in Fig.10 and Fig.12, is a clear difference in the time dependence
of the entropy $S(t)$. Comparing the data in the insets, one can conclude
that linear increase of the entropy occurs for the case when the form of the
SF is close to the Gaussian, but not for the case of the BW-form. Indeed, it
is hard to indicate clear time scale of a linear increase of $S(t)$ in
Fig.10, in contrast to Fig.12 where the linear time dependence is clearly
seen (apart from a very small time scale). This very fact may be of quite
generic, since in the TBRI model we also see a non-linear character of the
entropy increase in the BW-regime.

\section{Acknowledgements}

This work was supported by the Australian Research Council.  One of us
(F.M.I.) gratefully acknowledges the support by CONACyT (Mexico) Grant No.
34668-E.

\end{multicols}
\end{document}